\def \simlt{\lower.5ex\hbox{$\; \buildrel < \over \sim \;$}}
\def \simgt{\lower.5ex\hbox{$\; \buildrel > \over \sim \;$}}

\documentclass{elsart}

\usepackage{graphicx}

\begin{document}

\begin{frontmatter}

\title{Mapping the CMB sky: THE BOOMERANG experiment}

\author{P. de Bernardis$^1$, P.A.R. Ade$^2$, R. Artusa$^3$, J.J. Bock$^3$, A. Boscaleri$^4$, } 
\author{B.P. Crill$^3$, G. De Troia$^1$, P.C. Farese$^5$, M. Giacometti$^1$, V.V. Hristov$^3$,} 
\author{A. Iacoangeli$^1$, A.E. Lange$^3$, A.T. Lee$^6$, S. Masi$^1$, L. Martinis$^7$,}
\author{P.V. Mason$^3$, P.D. Mauskopf$^8$, F. Melchiorri$^1$, L. Miglio$^1$, T. Montroy$^5$, }
\author{C.B. Netterfield$^3$, E. Pascale$^{3,4}$, F. Piacentini$^1$, P.L. Richards$^6$, J.E. Ruhl$^5$, }
\author{F. Scaramuzzi$^7$}

\address{$^1$ Dipartimento di Fisica, Universita' La Sapienza, Roma, Italy}
\address{$^2$ Queen Mary and Westerfield College, London, UK}
\address{$^3$ California Institute of Technology, Pasadena, CA, USA}
\address{$^4$ IROE-CNR, Firenze, Italy}
\address{$^5$ Dept. of Physics, Univ. of California, Santa Barbara}
\address{$^6$ Dept. of Physics, Univ. of California, Berkeley}
\address{$^7$ ENEA CRE Frascati, Italy}
\address{$^8$ Dept. of Physics and Astronomy, University of Massachussets}

\begin{abstract}
We describe the BOOMERanG experiment, a stratospheric balloon
telescope intended to measure the Cosmic Microwave Background
anisotropy at angular scales between a few degrees and ten
arcminutes. The experiment has been optimized for a long
duration (7 to 14 days) flight circumnavigating Antarctica at the end
of 1998. A test flight was performed on Aug.30, 1997 in Texas.
The level of performance achieved in the test flight was
satisfactory and compatible with the requirements for the long
duration flight.

\end{abstract}

\end{frontmatter}

% main text

\section{Introduction}

The anisotropy of the CMB at angular scales between
a few arcminutes and a few degrees is expected to provide
invaluable information on the structure and evolution
of our Universe (see e.g. Hu et al. 1996, 1998, 
Lineweaver et al. 1998, Tegmark 1998).
Two major satellite experiments are planned,
MAP in 2001 (NASA) and Planck Surveyor in 2006 (ESA),
each with full sky and wide frequency coverage.  For optimal
design of these missions smaller experiments are required,
which cover smaller sky areas with comparable sensitivity
and similar observation strategy, using the critical
technologies presently under development for satellite
missions. Several balloon-borne
experiments are currently being developed (see de Bernardis and Masi, 1998
for a review). Here we describe the BOOMERANG (Balloon
Observations Of Millimetric Extragalactic Radiation
ANisotropy and Geophysics) experiment, which was flown
for the first time on Aug.30, 1997. The experiment
is expected to measure the CMB anisotropy with
unprecedented accuracy, probing the theory
of evolution of structures in the Universe.
In addition, it will test detector and scan technologies
critical to Planck HFI.
The system has an angular resolution of 12 arcmin FWHM 
and is a scanning telescope, making wide (up to 70$^o$
peak to peak) fast (2$^o$/s) azimuth scans
while the sky drifts, thus achieving significant
sky coverage. It has been designed to be launched 
on a long duration balloon flight, circumnavigating
the Antarctic continent in Dec.1998. The sky region
opposed to the sun during the Antarctic summer 
is a dust-free zone, in the Horologium costellation. 
The diffuse dust emission
in this region ranges between 0.2 and a few MJy/sr 
in the IRAS maps at 100 $\mu m$  (Schlegel et al. 1997) 
with degree-scale gradients of the
order of 0.1 MJy/sr.  The brightness extrapolated
at 150 GHz assuming a single temperature dust model
($T_d = 18 K$, $\epsilon = \epsilon_o (\nu / \nu_o) ^{1.7}$)
is equivalent to temperature fluctuations of the 
CMB smaller than a few $\mu K$ at 150 GHz, making this
region a very appealing target for cosmological observations.
The significant advantage of these antarctic balloon
flights is the flight duration, which can exceed
10 days. Taking advantage of this, we can map a wide region
(45 deg x 25 deg, corresponding to 30000 independent
pixels at our resolution) with significant integration 
time per pixel (10 to 100 seconds) and deep repeated checks 
for systematic effects.

\section{\bf The Instrument}

The BOOMERANG experiment is a millimeter wave
telescope, with a bolometric receiver working
in a long-duration cryostat at 0.3 K for 15 days.
The instrument has been optimized for the peculiar
requirements of Antarctic long duration ballooning (LDB).
A diagram of the instrument is presented in fig.1.

\begin{figure}[!h]
\begin{center}
\includegraphics[height=10cm]{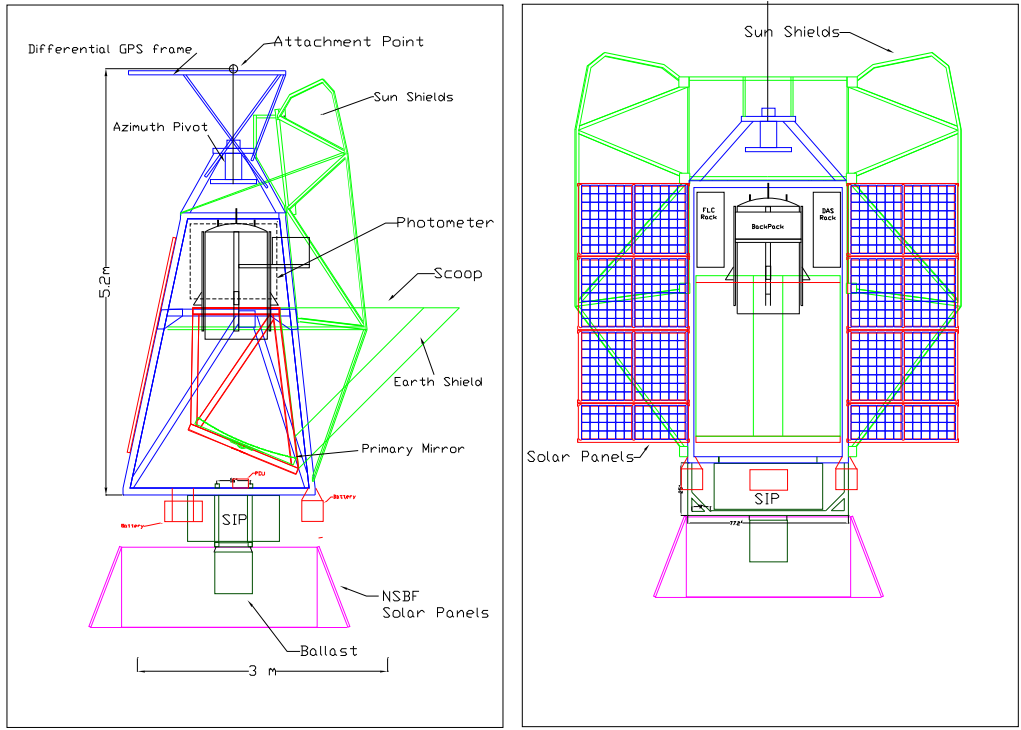}
\caption{The BOOMERanG Payload in the LDB configuration}
\end{center}
\end{figure}

The primary mirror of the telescope (a 1.3 m 
diameter, 1.5 m focal length, off-axis
paraboloid) is at ambient temperature. It is
made out of 6061 aluminum, as is the entire telescope
frame. At the focus of this telescope, the multiband receiver 
features beams ranging from 12 to 20 arcmin FWHM
(depending on the channel and on the configuration).
The telescope is protected by an earth shield
and by two large sun shields (see fig.2), which allow operation
of the system in the range $\pm 60^o$ from the
anti-solar azimuth. Off axis radiation at 150 GHz
is undetected at the -85dB level.

\begin{figure}[!h]
\begin{center}
\caption{The BOOMERanG payload hanging from the launch
vehicle on Aug.30, 1997. The earth shield and the 
two sun shields dominate the scene.}
\end{center}
\end{figure}

The secondary and tertiary mirrors are cooled
to 2K inside the cryostat. Radiation enters the
cryostat through a 50 $\mu m$ thick polypropylene
window, and goes through blocking filters at 77K
and at 4.2K. The tertiary mirror
is the cold Lyot stop of the instrument, sharply
limiting the field of view of the photometers.
Throat-to-throat concentrators with their entrance
aperture placed in the telescope focal plane
allow radiation to enter inside the RF-tight box 
containing the $^3$He refrigerator and the
detectors. 
The detectors are based on parabolic and conical
concentrators, metal mesh filters and Si$_3$N$_4$
spider web absorber bolometers. 
A photograph of the focal plane array for the Antarctic
flight and of its ancillary hardware is shown in fig.3.

\begin{figure}[!h]
\begin{center}
\caption{The cryogenic insert for the BOOMERanG 1998 receiver.
The gold plated flange is supported by vespel columns and is
cooled to 0.28K by the $^3$He fridge visible in the background.
Above the flange the multiband photometers and the
related wiring are visible. Below the flange is visible
the cryogenic preamplifier box (lid open). The insert shown
in the picture is 60 cm high and is bolted on the 2K flange of the 
main long duration dewar.
}
\end{center}
\end{figure}

We have two different configurations for the focal plane.
For the test flight we traded angular resolution for throughput
in order to get significant sensitivity during the short flight from
Texas (6 hours). The final configuration, used for the Antarctic
flight, has higher angular resolution and improved detectors.
The details are summarized in fig.4. The two configurations
have been designed to allow several levels of signal confirmation
and detector redundancy during the scan (see Masi et al. 1997 for 
details).

\begin{figure}[!h]
\begin{center}
\includegraphics[height=10cm]{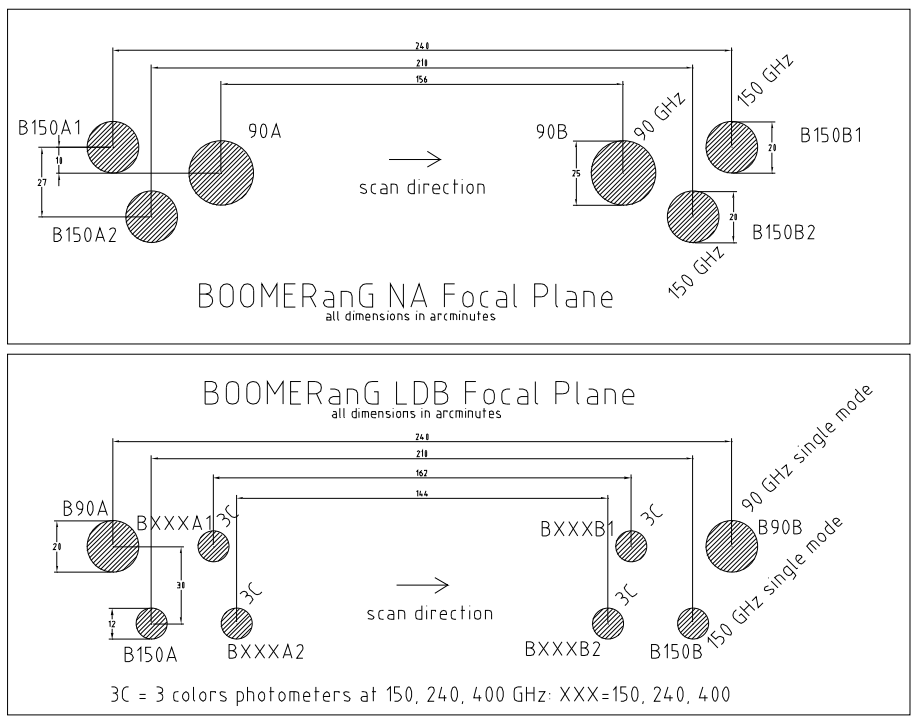}
\caption{ The two focal plane configurations of the
BOOMERanG receiver. The band centers and widths are as follows
($\nu_o, \Delta \nu FWHM$), in GHz:
B90X=(90,25), B150X=(160, 41), B220X=(235, 50), B400X=(412, 32).}
\end{center}
\end{figure}

Low background bolometers are extremely sensitive to
all forms of radiant energy. One big problem with standard
bolometers is energy deposition from cosmic rays (CR) particles.
Previous experiments carried out in temperate regions
measured typical rates of an event every few seconds.
The problem is enhanced in the polar stratosphere, where we 
expect a flux of cosmic rays about 6 times larger than in the
temperate stratosphere. Bolometers with negligible
cross sections for cosmic rays have been developed
(Bock et al. 1995, Mauskopf et al. 1997) and tested
in flight. The absorber is a thin web 
($\sim 1 \mu m$ thickness)
micromachined from Si$_3$N$_4$ and metalized. 
The grid constant is of the order of a few hundred microns, 
smaller than the in band wavelength so that
photons are effectively absorbed while CR are not.
%These bolometers have photon absorption efficiency similar
%to standard composite bolometers with solid absorbers, while
%the cross section for CR is reduced to a few per cent.
Also the heat capacity of the bolometer is greatly reduced,
and the mechanical resonant frequency is in kHz range.
NEPs of the order of $10^{-17} W/\sqrt{Hz}$ at 0.3K are achieved
with time constants of the order of 10 ms for the 150GHz
detectors. The other form of energy which contaminates
signals from high sensitivity bolometers is radio frequency
(RF) pickup, either through the optical path or 
through the readout circuit.
This problem is very severe in the case of balloon borne
payloads, which have powerful GHz telemetry transmitters on
board. The BOOMERanG receiver has been carefully shielded: all the
detectors work in a RF tight cavity. Millimeter-wave radiation enters the 
cavity through apertures much smaller than the RF wavelength 
(inside the throat to throat parabolic concentrators), 
and all the wires enter the cavity through cryogenic EMI filter 
feedthroughs. Similar care has been taken for the signal 
processing circuits.

The bolometers are differentially AC biased and read-out;
the bolometer signals are demodulated by lock-ins synchronous
with the bias voltages. A RC cut-off at a frequency (15 mHz) lower
than the scan frequency was used to limit the dynamical range of
the data and to remove 1/f noise from very low frequencies,
unimportant for our measurements.

A long duration $^4He$ cryostat (Masi et al. 1998B) 
cooling the experimental insert to 2K, and a long
duration $^3He$ fridge (Masi et al. 1998A), cooling 
the photometers to 280 mK, have been developed.
The cryostat has a central 50 liter volume available
for the 2K hardware, including blocking filters,
secondary and tertiary mirrors, cryogenic preamplifiers,
back to back concentrators, $^3$He fridge and photometers.
The cryogens hold time exceeds 12 days under flight conditions. 

The telescope and receiver hardware are mounted on a tiltable
frame (the inner frame of the payload). The observed elevation
can be selected by tilting the inner frame by means of a 
linear actuator. The outer frame of the payload is connected 
to the flight train through an azimuth pivot. The observed azimuth
can be selected by rotating the full gondola around the pivot,
by means of two torque motors. The first actuates a flywheel,
while the second torques directly against the flight train.
An oil damper fights against pendulations induced 
by stratospheric wind shear or other perturbations.
The sensors for attitude control are different for night-time
flights (like the test flight in 1997) than day-time flights
(like the Antarctic flight in 1998). For the night-time flight
in Texas we had a set of 3 vibrating gyroscopes and a flux-gate
magnetometer in the feedback loop driving the sky scan. 
A CCD camera with real-time star position measurement
 provides attitude reconstruction to 1 arcmin.
For the day-time flight in Antarctica we have a coarse and a fine
sun sensor, a differential GPS, and a set of 3 laser gyroscopes.
Again, we expect to be able to reconstruct the attitude
better than 1 arcmin.

\section{\bf Observations}

The system was flown for 6 hours on Aug 30, 1997, from the
National Scientific Balloon Facility in Palestine, Texas.
All the subsystems performed well during the flight:
the He vent valve was opened at float and was closed at termination,
the Nitrogen bath was pressurized to 1000 mbar,
the $^3$He fridge temperature (290mK) drifted with 
the $^4$He temperature by less than 6mK during the flight.
Pendulations were not generated during CMB scans, at a
level greater than 0.5 arcmin, and both azimuth scans 
at 2$^o$/s and full azimuth rotations (at 2 and 3 rpm) 
of the payload were performed effectively. 
The loading on the bolometers was as expected,
and the bolometers were effectively CR immune, with
white noise ranging between 200 and 300 $\mu K \sqrt{s}$. 
During the flight, the system observed
Jupiter to measure the beam pattern and responsivity. 
In fig.5 we plot an example of signal from scans on Jupiter.
\begin{figure}[!h]
\begin{center}
\includegraphics[height=9cm, width=11 cm]{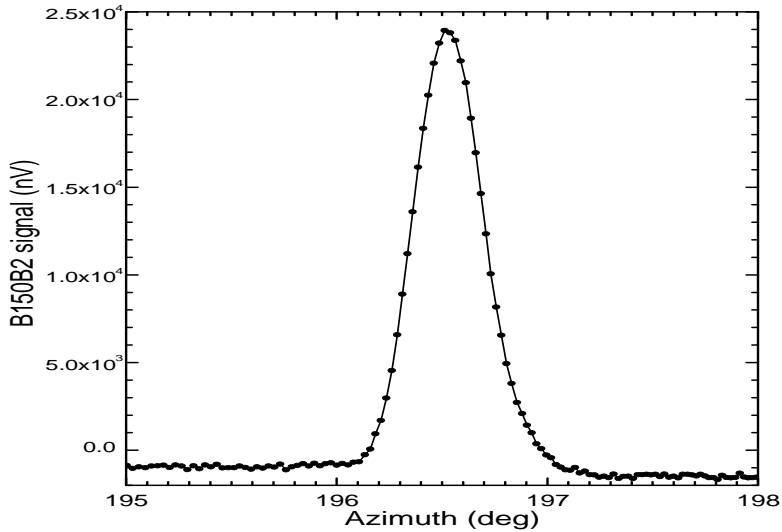}
\caption{ Sample real time signal from a 150 Ghz bolometer
during azimuth scans on Jupiter. The scan speed is 1.7$^o/s$
towards increasing azimuth,
the elevation of the beam is 39$^o$. Due to the non-negligible time
constant of the bolometer, the signal appears non-symmetric.
The effect of the high-pass filter is also evident in the
baseline. These effects are removed by deconvolution from 
system transfer function in the data analysis.}
\end{center}
\end{figure}
We also performed 360$^o$ azimuth spins of the payload,
to measure the Dipole anisotropy of the CMB for an independent
estimate of the responsivity. In fig.6 we plot an example 
of real-time data during the spin, showing the
CMB dipole signature. 
\begin{figure}[!h]
\begin{center}
\includegraphics[height=9cm, width=11 cm]{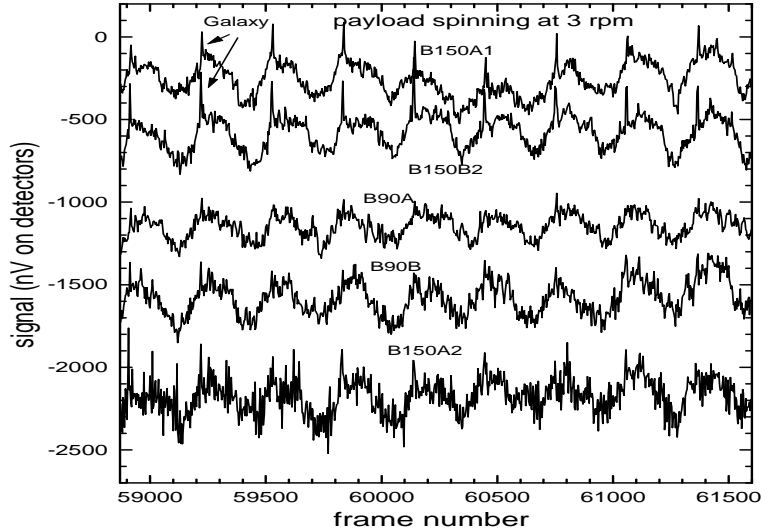}
\caption{ Sample section of data taken during fast payload spin
(3 rpm). The dipole anisotropy of the CMB is evident in all the
active channels
}
\end{center}
\end{figure}
A strip of sky at high galactic latitudes ($\sim$ 4 deg high 
in declination, $\sim$ 6 hours wide in RA) in the constellations 
of Sagittarius, Capricornus, Aquarius and Sculptor, was scanned 
in search for CMB anisotropies. The gondola performed
40$^o$ peak to peak azimuth scans, centerd on the south, 
with 50 s period, for 5 hours.

\section{\bf Analysis}

The first step in telescope attitude reconstruction was
the measurement of the offset between the microwave telescope 
beams and the CCD camera during Jupiter observations. 
Second, we used the data from the magnetometer
and the elevation encoder as a zero order solution for the attitude, 
and found precise solutions for the pointing using the star positions
measured by the CCD camera, and interpolationg between solutions
using the gyroscope data.  
The temporal transfer function of all the channels
was measured using Cosmic Rays Pulses (a total of 60 in 6 hours
in the NA-B150B2 channel) and fast scans on Jupiter during
the dipole rotations. The raw data were deconvolved for the
temporal transfer function and deglithced.
Time ordered data were spectrally analyzed searching for
bias aliases and interference lines. 
Data sections contaminated were FT filtered to remove the
lines. Data sectors contaminated by obvious spurious
effects were removed (cosmic rays hits, calibration lamp spikes,
microphonic events from nitrogen valve, Radar Hits, 
Jupiter scans).  The resulting "clean" time ordered
data set is mainly noise, and is stationary: the rms 
changes less than 2$\%$ from hour to hour. 

The photometer responsivities were 
measured using the scans on Jupiter and
the spectral response measured before
the flight by means of a fourier transform interferometer.
Data were filtered through a flat-phase 
numeric filter to reduce white noise. Simulations were made 
to select the best parameters for the filter. Filtered data 
were binned on a grid centered on the optical position of Jupiter.
The beam pattern, $\Omega _{BEAM}$ and FWHM were computed from 
the binned data (see table 1). 
The signal was estimated making fits to the 
scan data. The responsivity was derived using the measured 
quantitites and the spectral response measured pre-flight.
\begin{table}
\begin{center}
\begin{tabular}{|l|r|c|} \hline
channel         &         FWHM       &       $\Omega(sr)$         \\ \hline
NA-B150A1          &    $19.5^\prime$   & $(3.04 \pm 0.17)10^{-5}$   \\ \hline
NA-B150B1          &    $19^\prime$     & $(3.03 \pm 0.25)10^{-5}$   \\ \hline
NA-B150B2$(1^{st})$&    $16.5^\prime$   & $(2.32 \pm 0.09)10^{-5}$   \\ \hline
NA-B150B2$(2^{nd})$&    $16.5^\prime$   & $(2.31 \pm 0.10)10^{-5}$   \\ \hline
NA-B90A            &    $24^\prime$     & $(5.61 \pm 0.36)10^{-5}$   \\ \hline
NA-B90B            &    $29^\prime$     & $(5.68 \pm 0.27)10^{-5}$   \\ \hline
\end{tabular}
\end{center}
\caption{\textit{Beam Size measurements for the BOOMERanG NA telescope}}
\end{table}
An independent measurement of the photometer's responsivity
was done using the dipole signal, which has the same spectrum
as smaller scale CMB anisotropies and is well known. 
On our scans (obtained spinning the gondola at 2 and 3 rpm) 
the expected dipole signal is between 2 mK and 4 mK peak to  peak.
We do see a dipole like signal both at 90 GHz and at 150 GHz.
Atmospheric effects cannot produce the observed signal.
Repeating the measurements four times at 1 hour intervals, 
we check that the observed dipole rotates with the sky,
and has the expected pattern. From the amplitude of the
observed signal we get a responsivity independent on the
spectral response of the instrument. The responsivities measured
at the same time during the flight with the two independent methods
are consistent within the error bars ($\simlt 5 \%$) at 
one $\sigma$. An internal stimulator lamp, placed in 
the Lyot stop of the optical system and flashing every
15 minutes, is used to transfer the calibration to the
subsequent parts of the flight. 

The CMB data analysis is in progress, along two parallel
paths. In the first one the data are binned in synthesized
patterns as in Netterfield et al. (1996). The
corresponding window functions are computed, and a 
likelihood analysis of the binned data is performed.
Band-power estimates for the $c_\ell$-s are computed.
In the second approach, a map is created from the data
(see e.g. Janssenet al. 1996, Wright 1996, Smoot 1997,
Tegmark 1997). The BOOMERanG test flight data produce a 
26000 pixels map  ($6^\prime \times 6^\prime$ pixels). 
The map is big, as a result of the only scan strategy 
compatible with our late flight (Aug.30).
At the pixel level, the map is dominated by noise. 
The spherical harmonics analysis of the map is a significant
computational problem, and is in progress.
The analysis is done for all the channels and 
for the dark channel, in order to make correlation 
analysis and spectral checks. 
Both the data analysis processes are being iterated
with different data selection rules, 
deconvolution parameters, noise estimates
to check for systematic effects.

The observed region contains slightly dusty sections and 
very foreground-clean sections.
Both are interesting. Cross correlations of the
BOOMERanG map with the 
DIRBE/IRAS dust maps by Schlegel et al. (1997)
will provide important insight on the topic of
mm-wave emission of interstellar dust.

A careful analysis of the possible systematic effects is still in 
progress. At present we can say only that anisotropy is evident 
in the measured data, and is measured with good signal to noise 
ratio in the range $100 < \ell < 400$.

%\begin{table}[!h]
%\begin{center}
%\begin{tabular}{|c|c|c|}
%\hline
%
%$\nu$  & $\Delta \nu$    & $\Delta T_{CMB}$         \\
%$GHz$  & $GHz$           & $mK$                     \\
%\hline
%40     &  20           &   $1.65 \times 10^{-4}$    \\
%90    &  20            &   $2.89 \times 10^{-3}$    \\
%150   &  36            &   $8.29 \times 10^{-3}$    \\
%474   &  54            &   $0.226              $    \\
%675   &  39            &   $0.248              $    \\

%\hline
%\end{tabular}
%\end{center}
%\caption{ CMB Temperature fluctuations $\Delta T_{CMB}$ 
%equivalent to stratospheric brightness fluctuations in different 
%frequency bands (band center $\nu$, FWHM $\Delta \nu$). }
%%\label{tabfital}
%\end{table}

\section{\bf Conclusions}

The BOOMERanG experiment has been successfully flown from Texas
in a short (6 hours) test flight. The flight has proven satisfactory 
performance of the instrument and significant science data. 
A 26000 pixel map, the largest ever produced in CMB experiments,
has been produced at 90 GHz and 150 GHz. The data analysis is
in progress. The payload
has been upgraded with a 16 detectors focal plane,
and the addition of the hardware necessary for a long duration flight
(solar panels for power supply and satellite communications).
The system is presently in Antarctica to be launched by NSBF
for the first long duration flight.

\section{\bf Acknowledgments}

The BOOMERanG experiment has been funded by 
Universita' di Roma La Sapienza, Agenzia Spaziale Italiana (ASI) and 
Programma Nazionale di Ricerche in Antartide (PNRA)
in Italy, NASA and Center for Particle Astrophysics in USA, 
PPARC in UK. PdB acknoledges support from the Consorzio Internazionale
di Fisica Spaziale during his stay at the International School
of Space Physics.

\end{document}